\documentclass[twocolumn,english]{revtex4}
\usepackage[T1]{fontenc}
\usepackage[latin9]{inputenc}
\usepackage{amssymb}
\usepackage{graphicx}

\makeatletter
\@ifundefined{textcolor}{}
{%
 \definecolor{BLACK}{gray}{0}
 \definecolor{WHITE}{gray}{1}
 \definecolor{RED}{rgb}{1,0,0}
 \definecolor{GREEN}{rgb}{0,1,0}
 \definecolor{BLUE}{rgb}{0,0,1}
 \definecolor{CYAN}{cmyk}{1,0,0,0}
 \definecolor{MAGENTA}{cmyk}{0,1,0,0}
 \definecolor{YELLOW}{cmyk}{0,0,1,0}
 }

\makeatother

\usepackage{babel}
\begin{document}

\title{Liquid-liquid phase transition model incorporating evidence for ferroelectric
state near the lambda-point anomaly in supercooled water. }

\author{P.O. Fedichev}

\affiliation{Quantum Pharmaceuticals Ltd, Ul. Kosmonavta Volkova 6a-1205, 125171,
Moscow, Russian Federation, http://q-pharm.com}

\email{peter.fedichev@gmail.com}

\author{L.I. Menshikov}

\affiliation{NRC Kurchatov Institute, Kurchatov Square 1, 123182, Moscow, Russian
Federation}
\begin{abstract}
We propose a unified model combining the first-order liquid-liquid
and the second-order ferroelectric phase transitions models and explaining
various features of the $\lambda$-point of liquid water within a
single theoretical framework. It becomes clear within the proposed
model that not only does the long-range dipole-dipole interaction
of water molecules yield a large value of dielectric constant $\epsilon$
at room temperatures, our analysis shows that the large dipole moment
of the water molecules also leads to a ferroelectric phase transition
at a temperature close to the $\lambda-$point. Our more refined model
suggests that the phase transition occurs only in the low density
component of the liquid and is the origin of the singularity of the
dielectric constant recently observed in experiments with supercooled
liquid water at temperature $T\approx233K$. This combined model agrees
well with nearly every available set of experiments and explains most
of the well-known and even recently obtained results of MD simulations.
\end{abstract}
\maketitle
Supercooled water exhibits a number of well-known anomalies near the
so-called $\lambda-$point at temperature $T_{\lambda}\approx228K$
(at normal pressure), where a good number of thermodynamic quantities
such as heat capacity, compressibility, thermal expansivity, and dielectric
constant all exhibit nearly singular behavior \cite{angell1973anomalous,speedy1976isothermal,hodge1978relative}.
The intrinsic thermodynamic instability of liquid water at temperatures
well below the freezing point has been a major obstacle both in experimental
studies and theoretical modeling (see e.g. \cite{mishima1998relationship,debenedetti2003supercooled,DebenedettiStanley2003,angell2004amorphous,stanley2010water,buldyrev2007water,han2010phase,kumar2009anomalies}
for a review). The observed features are often weak, which suggests
a thermodynamic continuity of the various water states near the $\lambda-$point
and relates the observed {}``singularity'' with essentially a random
sharp feature, a function of the parameters of the liquid \cite{sastry1996singularity,rebelo1998singularity}.
A considerably more accepted view is to ascribe the features characteristic
of a phase-transition to a first-order liquid-liquid phase transition
(LLPT). This model predicts the existence of a second critical point
of water at the temperature $T_{{\rm CR}}\sim200K$ and pressure $\sim1kbar$
\cite{poole1992phase,poole1994effect}. The sharp temperature dependencies
observed near $T_{\lambda}$ are attributed to crossing the Widom
line \cite{xu2005relation}, where the density and entropy fluctuations
are large \cite{bertrand2011peculiar} and which happens at $T\approx T_{\lambda}$
at normal pressure. The view is supported by numerous molecular dynamics
(MD) simulations based on realistic water models \cite{poole1992phase,stanley1994there,tanaka1996phase,harrington1997equation,yamada2002interplay,jedlovszky2005liquid,paschek2005liquid,paschek2008thermodynamic,liu2009low,abascal2010widom},
simplified analytical models \cite{truskett1999single,truskett2002simple,lee2001simple,girardi2007liquid,stokely2010effect},
and experimental studies \cite{mishima1998relationship,mishima2000liquid,mallamace2008nmr,mishima2010volume}.
The reported anomalies are not restricted to static features, the
dynamic properties such as the Einstein relation between diffusion
and mobility coefficients \cite{chen2006violation,kumar2007relation}
and the Arrhenius behavior of the liquid's dynamic properties \cite{xu2005relation,kumar2007relation,kumar2008predictions}
break down near $T_{\lambda}$ as well. 

Recent MD studies have demonstrated that heat capacity and thermal
conductivity \cite{kumar2011thermal} peak around $T_{\lambda}$ as
well and that the liquid shows a good deal of ordering in the vicinity
of the $\lambda-$point. Similarly, recent measurements of the dielectric
properties of liquid water confined in nanopores and hence prevented
from freezing well below the natural freezing point \cite{schreiber2001melting}
manifest a profound bump in the dielectric constant near the $\lambda-$point
\cite{fedichev2011experimental,bordonsky2011}. These observations
paint a richer picture than a mere first-order liquid-liquid phase
transition and, in fact, bring back an old idea \cite{hodge1978relative}
relating the weak singularity of the dielectric constant to a ferroelectric
phase transition (FPT). Remarkably, the hypothesis was put forward
immediately after the discovery of the $\lambda-$point, though the
weakness of the observed singularities prompted the authors \cite{hodge1978relative}
to reject the explanation. Furthermore, the ferroelectric instability
for a model liquid with parameters similar to water is predicted to
occur at a very high and essentially unreachable temperature, $\sim1200K$
\cite{bernal1933theory}. A more sophisticated model \cite{fedichev2008fep,fedichev2006long,men2011}
predicts a ferroelectric phase transition at a temperature independent
of the details of the short-range interactions between the molecules,
\begin{equation}
T_{F}=\frac{4\pi n_{0}d_{0}^{2}}{9\epsilon_{\infty}}=210\div236K\approx T_{\lambda},\label{eq:Tc}
\end{equation}
where $n_{0}$ is the density and $d_{0}$ are the static dipole moments
of the molecule comprising the liquid. The dielectric constant $\epsilon_{\infty}$
is not associated with the molecules' degrees of orientational freedom
and comes from electron shell polarization, $\epsilon_{\infty}=4\div5.5$
(as discussed in e.g. \cite{stogrin1971,liebe1991model,hasted1948dielectric}).
Despite bringing the phase transition temperature into the right range,
the model predicts a behavior of the dielectric constant with changes
in temperature that is far too gradual when compared with empirical
results \cite{fedichev2011experimental,bordonsky2011}. Consequently,
it cannot even qualitatively explain all of the features of supercooled
water by itself. Realistic liquids such as water are far more complicated
than a model polar liquid consisting of point dipoles. For example,
the tetrahedral geometry of $H_{2}O$ molecule and its electron shell
leads to a polyamorphism phenomenon \cite{poole1995amorphous,mishima1998relationship,angell2004amorphous},
namely two or more phases of the same liquid existing in a mixture
at the same time. 

To unify the observed ferroelectric-like properties of water-molecule
ordering and the singularity of the dielectric constant near $T_{\lambda}$
\cite{fedichev2011experimental,bordonsky2011} with the previously
reported signatures of the LLPT within a single theoretical framework,
we combine our simple polar liquid phenomenology \cite{fedichev2008fep,fedichev2006long,men2011,fedichev2011experimental}
with the LLPT hypothesis \cite{poole1992phase,poole1994effect,mishima1998relationship}
using a two-component mixture model of water \cite{ponyatovskii1994pis}.
We assume that the equilibrium state of supercooled water is a mixture
of macroscopically-sized clusters of the two types: low density (LDL)
and high density liquid (HDL).The LDL local lattice is softer than
that of HDL and the density of HDL exceeds that of LDL by $\sim20\%$
\cite{mishima1998relationship,myneni2002spectroscopic,wernet2004structure,tokushima2008high,huang2009inhomogeneous,english2011density}.
Since LDL is {}``softer'', the molecules of the LDL rotate more
or less freely, whereas in the HDL the rotations are more difficult.
This explains why HDL has no ferroelectric state at any temperature.
We assume that the ferroelectric ordering and the FPT apparently observed
in supercooled liquid bulk water occurs in the LDL component only. 

The Gibbs free energy of an LDL cluster at a given pressure $P$ is
the sum of the contributions from the polar liquid, $G_{LDL}^{P}$,
and the lattice, $G_{LDL}^{L}$: $G_{LDL}=G_{LDL}^{P}+G_{LDL}^{L}.$
At high temperatures, $T>T_{F}$, the equilibrium state of LDL corresponds
to the disordered paraelectric phase, whereas at lower temperatures,
$T<T_{F}$, LDL undergoes a second-order phase transition and enters
the long-range-ordered ferroelectric state. Near the phase transition,
where $\tau=(T-T_{F})/T_{F}\ll1$, the free energy of LDL takes the
form:
\begin{equation}
G_{LDL}\approx-D\tau^{2}\theta\left(-\tau\right)+G_{LDL}^{L}\left(T,P\right),\label{eq: The molar energy of LDL}
\end{equation}
where $D\sim V_{0}n_{0}^{2}d_{0}^{2}\sim150\, cal/mol$, and $V_{0}=N_{A}/n_{0}\approx22\, cm^{3}$
is the molar volume of LDL. Since the LDL density differs by no more
than $20\%$ from the total liquid density, we will not distinguish
between the LDL density and $n_{0}$. The second-order FPT in LDL
manifests itself as a singularity of the dielectric constant $\epsilon$
(see e.g. \cite{frohlich1949theory}): 
\begin{equation}
\epsilon=\epsilon_{\infty}\left(1+f\left(T\right)\right),\label{eq:epsilonT}
\end{equation}
where $f\left(T\right)=3T_{F}/(T-T_{F})$ at $T>T_{F}$ and $f\left(T\right)=3T_{F}/2(T_{F}-T)$
at $T<T_{F}$, which is a much weaker dependence than that observed
experimentally \cite{fedichev2011experimental}. The discrepancy is
apparently due to the very sharp temperature dependence of the LDL
fraction $c$ near $T\approx T_{\lambda}$. To see that, we follow
\cite{ponyatovskii1994pis} and formulate a two-liquid model representing
the Gibbs energy of water as the energy of a two-liquid mixture of
macroscopic clusters using a representation similar to that used in
the physics of binary alloys (see e.g. \cite{krivoglaz1958teoriya}):
\[
G\left(c\right)=cG_{{\rm LDL}}+\left(1-c\right)G_{{\rm HDL}}+Uc\left(1-c\right)+
\]
\begin{equation}
+RT\left[c\log c+\left(1-c\right)\log\left(1-c\right)\right].\label{eq: Free energy in two-level approximation}
\end{equation}
Here $G_{HDL}$ is the free energy of the HDL component, and the parameter
$U$ characterizes the {}``energy of mixing''. The equilibrium free
energy and the LDL fraction $c$ are found through minimization of
$G\left(c\right)$ over $c$, $G^{\prime}\left(c\right)\equiv\left[\partial G\left(c,P,T\right)/\partial c\right]_{P,T}=0$,
or
\begin{equation}
\triangle G\left(P,T\right)+U\left(1-2c\right)+RT\log\left(\frac{c}{1-c}\right)=0,\label{eq: Implicit equation for c}
\end{equation}
where $\triangle G\left(P,T\right)=G_{LDL}-G_{HDL}$. The temperature
of the second critical point in the model is $T_{CR}=U/\left(2R\right)$
\cite{ponyatovskii1994pis}.
\begin{figure}
\includegraphics[width=0.9\columnwidth]{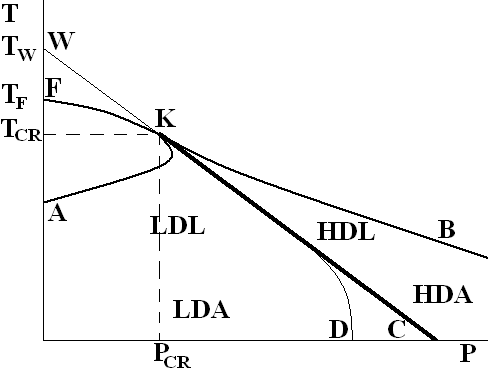}

\caption{Model PT diagram of water (see the explanations in the text). \label{fig: PT diagram of water}}
\end{figure}

To analyze these equations, we follow \cite{ponyatovskii1994pis,krivoglaz1958teoriya}
and assume that the thermodynamic quantities in $\triangle G\left(P,T\right)=\triangle E^{0}-T\triangle S^{0}+P\triangle V^{0}$
are practically temperature-independent:
\begin{equation}
\triangle E^{0}\left(P,T\right),\triangle S^{0}\left(P,T\right),\triangle V^{0}\left(P,T\right)\approx{\rm const}.\label{eq: Approximation from binary alloy theory}
\end{equation}
The PT diagram suggested by this model is shown in Fig.\ref{fig: PT diagram of water}.
The AKB line in the Figure is the spinodal line corresponding to $G^{\prime}\left(c\right)=0$,
$G^{\prime\prime}\left(c\right)\equiv\left[\partial^{2}G\left(c,P,T\right)/\partial c^{2}\right]_{P,T}=0$.
The KB line is the LDL spinodal, where the LDL phase loses its thermodynamic
stability (the local minimum of the function $G\left(c\right)$ corresponding
to the LDL phase disappears on this line). In turn, the KA line is
the HDL spinodal. The Gibbs potential $G\left(c\right)$ has a single
minimum everywhere above the AKB line and two minima below this line,
at $c=c_{1}$ and $c=c_{2}$. The section KC of the straight line
WC is the liquid-liquid first-order phase transition line corresponding
to the phase equilibrium conditions: $G\left(c_{1}\right)=G\left(c_{2}\right)$,
$G^{\prime}\left(c_{1}\right)=G^{\prime}\left(c_{2}\right)=0$. According
to Eqs.(\ref{eq: Free energy in two-level approximation}) and (\ref{eq: Implicit equation for c}),
these conditions are equivalent to $\triangle G\left(P,T\right)=0$,
which, according to approximation (\ref{eq: Approximation from binary alloy theory}),
means that $T$ is a linear function of $P$ on the KC line, which
means that the KC line is itself a straight line in Figure \ref{fig: PT diagram of water}.
The low and high density amorphous ice regions are denoted by LDA
and HDA.

\begin{figure}
\includegraphics[width=0.9\columnwidth]{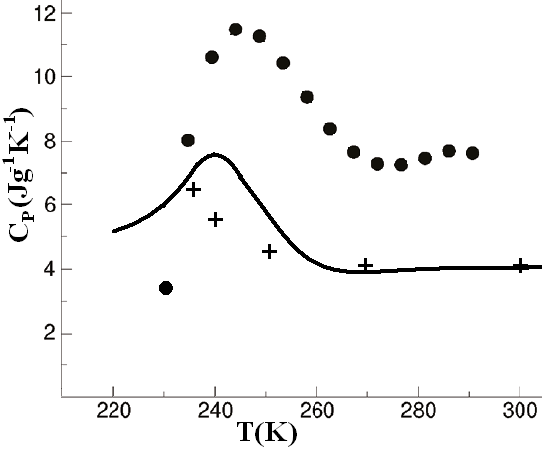}

\caption{Calculated water heat capacity (formulas (\ref{eq: Total heat capacity}),
(\ref{eq: C PP exact expression}), (\ref{eq: Parameters}); solid
curve) versus MD calculations \cite{kumar2011thermal} (points) and
empirical data \cite{angell1973anomalous} (crosses).\label{fig:Comparison-of-calculation} }
\end{figure}

The KD line in Figure \ref{fig: PT diagram of water} corresponds
to the limit $\triangle S\rightarrow0$ at $T\rightarrow0$ as required
by the third law of thermodynamics and which is never the case under
the assumptions (\ref{eq: Approximation from binary alloy theory}).
WK is the Widom line, defined by the conditions $G^{\prime}\left(c\right)=0$
and $G^{\prime\prime\prime}\left(c\right)=0$, or, equivalently, $c=1/2$.
At small pressures the Widom temperature is $T_{W}=\triangle E^{0}/\triangle S^{0}$.
The heat capacity, $C_{P},$ consists of the two parts

\begin{equation}
C_{P}=T\left[\left(\frac{\partial S}{\partial c}\right)_{P,T}\right]^{2}/G^{\prime\prime}\left(c\right)=C_{P}^{P}+C_{P}^{L},\label{eq: Total heat capacity}
\end{equation}
where $C_{P}^{L}$ is the lattice contribution, which can be calculated
e.g. using Debye approximation, $C_{P}^{L}=C_{P}^{\infty}\left(T/\Theta\right)^{3}/\left[1+\left(T/\Theta\right)^{3}\right]$,
where $C_{P}^{\infty}=18cal\cdot mol^{-1}K^{-1}$ (as suggested in
\cite{angell1973anomalous}) and $\Theta\approx150K$ is the Debye
frequency of water. Within the described model the polar part of $C_{P}^{P}$
is 
\begin{equation}
C_{P}^{P}\approx\frac{\left[\triangle E^{0}+P\triangle V^{0}+U\left(1-2c\right)\right]\left[\triangle S^{0}-R\log\left(\frac{c}{1-c}\right)\right]}{\frac{RT}{c\left(1-c\right)}-2U}.\label{eq: C PP exact expression}
\end{equation}
At small pressures along the Widom line $c\approx1/2$, $G^{\prime\prime}\left(c\right)$
is small, the fluctuations are strong, and the temperature dependence
of the heat capacity contribution
\begin{equation}
C_{P}^{P}\approx\frac{R\triangle^{2}}{\left[\left(T-T_{W}\right)^{2}+\delta^{2}\right]},\label{eq: Analytical form of peak dependence}
\end{equation}
takes a standard Lorenz form, where $\triangle=T_{W}-T_{CR}$, $\delta=2R\triangle\sqrt{\triangle/\left(\triangle E^{0}\triangle S^{0}\right)}$.
The quantity peaks at $T=T_{W}$ in agreement with experiments \cite{debenedetti2003supercooled,DebenedettiStanley2003},
earlier explanations \cite{xu2005relation}, and recent calculations
\cite{kumar2011thermal}. Therefore we can use Eq. (\ref{eq: Analytical form of peak dependence})
to analyze the heat capacity calculated e.g. using MD from \cite{kumar2011thermal}
and extract the model parameters: 
\begin{equation}
\triangle E^{0}=-860cal\cdot mol^{-1},\;\triangle S^{0}=-3.5cal\cdot mol^{-1}K^{-1},\label{eq: Parameters}
\end{equation}
and $U=880cal\cdot mol^{-1}.$ These parameters correspond to $T_{W}=245K$,
$T_{CR}=220K$, and $P_{CR}=1kbar$. To calculate $P_{CR}$ we used
$\triangle V^{0}\approx3.8cm^{3}mol^{-1}$ from \cite{ponyatovskii1994pis},
where very similar parameters were obtained: $U\approx900cal\cdot mol^{-1}$,
$\triangle E^{0}=-250cal\cdot mol^{-1}$, $\triangle S^{0}=-1cal\cdot mol^{-1}K^{-1}$,
and $T_{CR}=225K$. These {}``original'' parameters yield smaller
critical pressure value: $P_{CR}\approx0.33kbar$. The heat capacity
$C_{P}$ calculated with the help of Eqs. (\ref{eq: Total heat capacity}),
(\ref{eq: C PP exact expression}) and (\ref{eq: Parameters}) is
plotted on Fig.\ref{fig:Comparison-of-calculation} against the experimental
values from \cite{angell1973anomalous} and the recent MD calculations
\cite{kumar2011thermal}. Note that although the MD calculation does
not provide a full match with the experimental curve in absolute terms,
both data sets consistently describe the same feature and hence apparently
the same physics. This in combination with the water molecules ordering
predicted by the model (the entropy of LDL is less than that of HDL,
$\triangle S^{0}<0$, in accordance with \cite{DebenedettiStanley2003})
and confirmed by the MD calculations is the indication of ferroelectric
transition implicitly present already in the simulation \cite{kumar2011thermal}.

\begin{figure}
\includegraphics[width=0.9\columnwidth]{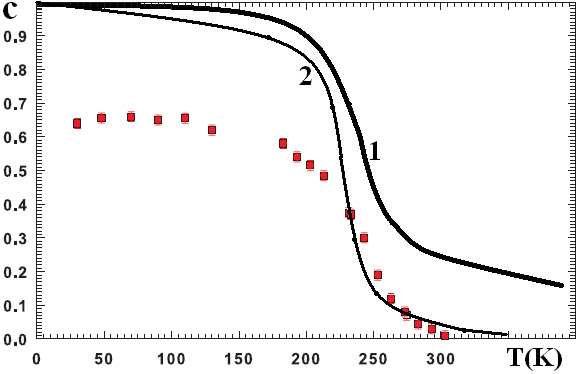}

\caption{Theoretical temperature dependence of the LDL fraction (solid lines,
see the explanations in the text) versus empirical data \cite{mallamace2008nmr}
(dots). \label{fig: Theory versus empirical data for C} }
\end{figure}

Once the model parameters (\ref{eq: Parameters}) are established
we can verify the consistency of the model by observing the temperature
dependence of the LDL fraction $c\left(T\right)$ given by Eq. (\ref{eq: Implicit equation for c})
(see solid line $1$ in Figure \ref{fig: Theory versus empirical data for C})
and compare it with the empirical data \cite{mallamace2008nmr} (the
dots in the same Figure). There is a qualitative agreement at least
at sufficiently large temperatures above $200K$. We note, that $\triangle G\left(P,T=0\right)<0$
as $T\rightarrow0$ and therefore, at sufficiently low temperatures
all of the liquid should turn into LDL, i.e. $c\rightarrow1$ as $T\rightarrow0$.
This leads us to believe that the equilibrium composition of water
was not actually achieved in the experiments of \cite{mallamace2008nmr},
even at very low temperatures. This may well be due to the increase
of equilibration time at extremely low temperatures, as discussed
in e.g. \cite{moore2010ice,moore2011structural}. With this in mind,
we can attempt to match the LDL fraction measurements only at higher
temperatures and obtain better agreement with the experiment using
slightly different model parameters 
\begin{equation}
\triangle E^{0}=-920cal\cdot mol^{-1},\;\triangle S^{0}=-4cal\cdot mol^{-1}K^{-1},\label{eq: Parameters 1}
\end{equation}
and $U=900cal\cdot mol^{-1}$ from \cite{ponyatovskii1994pis} (see
solid line $2$ in Figure \ref{fig: Theory versus empirical data for C}).
Although curves $1$ and $2$ are clearly different, both sets, (\ref{eq: Parameters})
and (\ref{eq: Parameters 1}), are very close numerically, which indicates
further difficulties in refining the parameters due to a very sharp
dependence of the LDL fraction $c$ next to the Widom line and clear
experimental difficulties. 

\begin{figure}
\includegraphics[width=0.9\columnwidth]{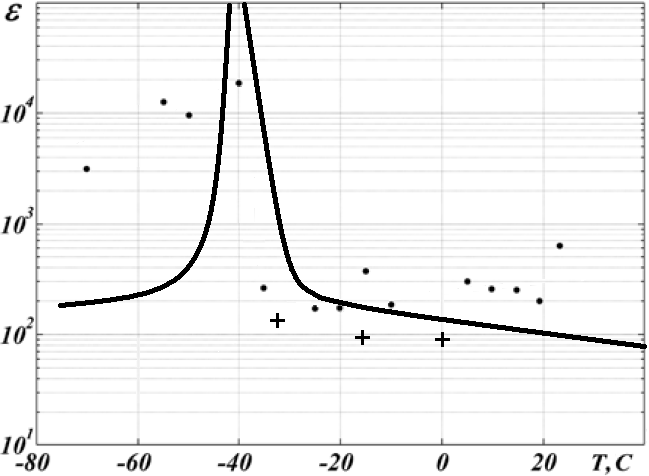}

\caption{Dielectric constant for bulk water versus temperature. Results of
formulas (\ref{eq: Final expression for dielectric constant}), (\ref{eq: Parameters})
are given by solid curve. Experimental results \cite{fedichev2011experimental,bordonsky2011}
are presented by points, crosses correspond to the measurements \cite{hodge1978relative}.\label{fig: Dielectric constant of water. Theory versus experiment.} }
\end{figure}

An analysis of the heat capacity and the LDL fraction measurement
let us verify the model and determine the model parameters. Consequently,
we may use the model to predict the temperature dependence of the
dielectric constant and compare it with empirical data. Accordingly,
the dielectric constant of the liquid is the sum of the LDL, $\epsilon_{LDL}=\epsilon_{\infty}(1+f(T))$
with $f(T)$ from Eq. (\ref{eq:epsilonT}), and the HDL contributions.
The HDL fraction is $1-c$, and its dielectric constant can be described
by the Debye-Onsager model \cite{Onsager}: $\epsilon_{HDL}=\epsilon_{\infty}+2\pi n_{0}d_{0}^{2}\left(\epsilon_{\infty}+2\right)^{2}/(9T)$.
Therefore the full expression $\epsilon=c\epsilon_{LDL}+(1-c)\epsilon_{HDL}$
becomes 
\begin{equation}
\epsilon=\epsilon_{\infty}\left[1+c\left(T\right)f\left(\tau\right)\right]+\left[1-c\left(T\right)\right]\frac{2\pi n_{0}d_{0}^{2}}{9T}\left(\epsilon_{\infty}+2\right)^{2}.\label{eq: Final expression for dielectric constant}
\end{equation}
The predicted temperature dependence calculated using the parameters
(\ref{eq: Parameters}) and the value $\epsilon_{\infty}=4.7$ corresponding
to $T_{F}=233K$ is compared with recent experimental measurements
\cite{fedichev2011experimental,bordonsky2011,hodge1978relative} in
Figure \ref{fig: Dielectric constant of water. Theory versus experiment.}.
At high temperatures, $T>T_{F},T_{W}$ the model dependence is not
far from the experimentally observed values. The temperature dependence
is now sharp, much stronger than that predicted by a simpler, single-component
model \cite{fedichev2011experimental}. This means that including
more liquid states in the analysis clearly improves the agreement
with experimental results. Below the transition point to the left
of the peak, at $T<T_{F}$, experiments yield distinctly different
results. There may be many reasons for that. For example it is quite
possible there is more than one LDL state of water below $T_{F}$.
The temperature $T_{F}$ itself may depend on the LDL cluster size,
and therefore we may face a {}``continuous set'' of ferroelectric
transitions in a multitude of LDL forms as the temperature decreases. 

When compared with available empirical data and numerical calculations,
the model calculations of the heat capacity, LDL fraction, and the
dielectric constant support a broader view, implying that supercooled
water is indeed a mixture of at least two different components, namely
LDL and HDL. There is a growing body of evidence for ferroelectric
transition in LDL clusters. Independent of the theoretical arguments
given above, the FPT in LDL can be supported by the structural similarity
between the local structure of LDL and the crystalline lattice of
stable ice Ih \cite{blackman1957cubic,finney2002structures}. At normal
pressure the paraelectric ice Ih enters the ferroelectric state, ice
XI, at temperature $T=72K$ \cite{kawada1972dielectric,jackson1997thermally,singer2005hydrogen,knight2006hydrogen}.
The ferroelectric state, ice XI, is obtained from hexagonal ice Ih
after reconstruction of the crystalline lattice, namely, by the shifting
of water molecules in each elementary cell of the crystal. It is interesting
to note that theory \cite{barkema1993properties} predicts antiferroelectric
ordering of molecular dipole moments for Ih ice with undeformed crystalline
lattice, but the ferroelectric state with a deformed lattice has a
lower free energy. It is not surprising that, due to a less dense
local crystalline lattice of LDL, the transition temperature observed
in \cite{fedichev2011experimental,bordonsky2011} essentially exceeds
the corresponding value of FPT {}``ice Ih$\rightleftharpoons$ice
XI''. It is also worth mentioning here that there have been important
studies concerning the ferroelectric states of metastable ice Ic,
with a cubic lattice first predicted in \cite{stillinger1977theoretical}
and detected experimentally in \cite{su1998surface,iedema1998ferroelectricity}.
The reported explanation of the {}``ice Ih$\rightleftharpoons$ice
XI'' transition gives us cause to believe that the ferroelectric
phase transition in LDL is also accompanied by the reconstruction
of the local crystalline lattice, i.e. the dipolemoment orientation
of molecules is strongly linked to the lattice's degrees of freedom.
On the other hand, FPT occurs only in LDL. Therefore, it is reasonable
to assume that the KC line in Figure \ref{fig: PT diagram of water}
is in fact the boundary between the ferroelectric and paraelectric
regions of the supercooled water. Here $\left|\triangle G_{P}\right|\sim\left|\triangle G_{L}\right|$
and therefore, the molecular orientation and the lattice's degrees
of freedom are equally important {}``driving forces'' of both the
ferroelectric and the liquid-liquid phase transitions. Above the FK
line the situations is different; there typically $\left|\tau\right|\ll1$
and Eq. (\ref{eq: The molar energy of LDL}) yields $\left|\triangle G_{P}\right|\ll\left|\triangle G_{L}\right|$.
Therefore, the lattice degrees of freedom dominate. Hence, we believe
that supercooled water is ferroelectric below the FKP line. 

\bibliographystyle{apsrev4-1}
\bibliography{../Qrefs}

\end{document}